\documentstyle[prl,aps]{revtex}
\input psfig
\def\fr#1/#2{{\textstyle{#1\over#2}}} 
\def\Uc{\hbox{\ss U}}
\def\>{\rangle}
\def\<{\langle}
\def\k#1{|\,#1\>}
\def\b#1{\<#1|}
\def\+{\oplus}
\def\ra{\rightarrow}
\font\ss = cmssbx10
\def\Hc{\hbox{\ss H}}

\def\Fc{\hbox{\ss F}}
\def\Fci{\hbox{\ss F}^\dagger}
\def\Pc{\hbox{\ss P}}
\def\Xc{\hbox{\ss X}}
\def\Zc{\hbox{\ss Z}}
\def\cX{\hbox{\ss cX}}
\def\cZ{\hbox{\ss cZ}}
\def\cXi{\hbox{\ss cX}^\dagger}
\def\cZi{\hbox{\ss cZ}^\dagger}
\def\oc{\hbox{\ss 1}}
\def\c{\fr1/{\sqrt2}}
\def\x{\otimes}

\begin{document}

\draft
\twocolumn[\hsize\textwidth\columnwidth\hsize\csname @twocolumnfalse\endcsname
\title{From~Classical~State-Swapping~to~Quantum~Teleportation}
\author{N.\ David Mermin}
\address{Laboratory of Atomic and Solid State Physics, 
Cornell University,  Ithaca, NY 14853-2501}
\maketitle

\begin{abstract} The quantum teleportation protocol is extracted
directly out of a standard classical circuit that exchanges the states
of two qubits using only controlled-NOT gates.  This construction of
teleportation from a classically transparent circuit generalizes
straightforwardly to $d$-state systems.\end{abstract} \pacs{PACS
numbers: 03.67.Hk, 03.67.Lx} ]

Quantum teleportation\cite{ft:teleportation} transfers the quantum
state of a two-state system (Alice's qubit, the source) to another
remote two-state system (Bob's qubit, the destination) without any
direct dynamical coupling between the two qubits.  To do this trick
Alice, who in general does not herself know the form of the state to
be transferred, must possess a third qubit (the ancilla) which
initially is maximally entangled with Bob's qubit in the two-qubit
state
\begin{equation}\c\bigl(\k0\k0+\k1\k1\bigr).\label{eq:epr}\end{equation}
Depending on the outcomes of appropriate measurements on the source
and ancilla, Alice can send Bob instructions that enable him to
transform the state of the destination into that originally possessed
by the source.  The term ``teleportation'' is apt because the
measurements that provide the information to recreate the state at the
destination obliterate all traces of it from the source.

If two qubits are allowed to interact, however, then their states can
be exchanged in a much less subtle way, with the help of three
controlled-NOT gates\cite{ft:cnot}.  The action of these gates can be
understood in entirely classical terms.  This is illustrated in
Fig.~1.

\begin{figure}
\centerline{\psfig{file=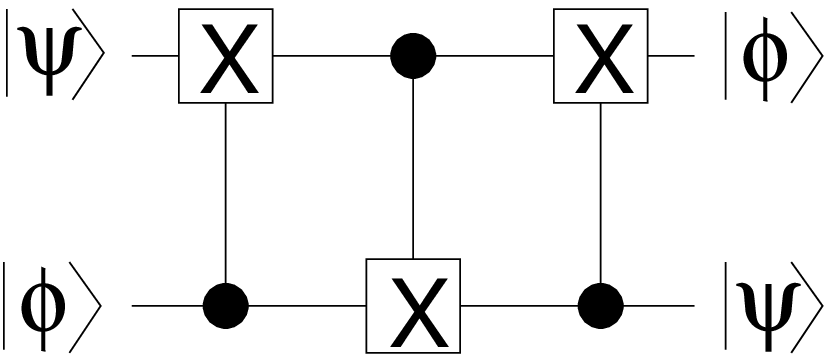,width=1.8truein}}
\caption{}
\end{figure}

\noindent That the classical~\cite{ft:classical} circuit in Fig.~1
does indeed exchange states is readily confirmed by letting it act on
a general computational basis state $\k x\k y$.  If $x$ is the value
(0 or 1) of the control bit and $y$ is the value of the target bit,
then the action of a single cNOT can be compactly summarized as
\begin{equation} \k x\k y \ra \k x\k{y\+x} \label{eq:cnot}
\end{equation} where $\+$ denotes addition modulo 2.  If $\k\psi = \k
x$ and $\k\phi = \k y$, then the action of the three successive gates
in Fig.~1 is (reading the Figure from left to right) \begin{equation}
\k x\k y \ra \k{x\+y}\k y \ra \k{x\+y}\k x \ra \k y\k x.
\label{eq:swap} \end{equation} This process makes perfect sense for
classical bits, as well as for quantum superpositions of classical
bits, to which it extends by linearity.
 
If the state $\k\phi$ in Fig.~1 is taken to be $\k
0$, then the cNOT gate on the left acts as the identity, so the
classical state-swapping circuit simplifies to: 

\begin{figure}
\centerline{\psfig{file=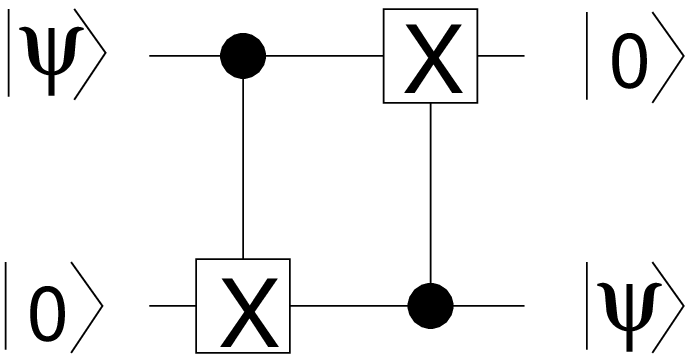,width=1.7truein}} 
\caption{} 
\end{figure} 

\noindent If the upper qubit (source) in Fig.~2 belongs to Alice and
the lower qubit (destination), to Bob, then this special case of the
general classical state-swapping circuit provides a considerably
simpler version of what happens in quantum teleportation.  But the
classical circuit in Fig.~2 is not teleportation,
because it requires direct dynamical couplings between the qubits
--- couplings that teleportation manages to avoid by the use of an
entangled pair of qubits and the classical communication of quantum
measurement outcomes.

In this note I illuminate the way in which quantum mechanics obviates
the need for the direct dynamical couplings in Fig.~2, showing
explicitly how this intuitive classical state-swapping circuit leads
directly to the transference of a state between uncoupled qubits that
constitutes quantum teleportation.  It is possible to eliminate all
direct couplings between the source and the destination because
quantum qubits have a richer range of logical capabilities than do
classical bits.  Only one indirect dynamical coupling between Alice
and Bob survives this process of elimination as the initial
interaction necessary to entangle Alice's ancilla with the Bob's
destination qubit.  All other direct dynamical coupling is replaced by
classical communication.

The key to relating quantum teleportation to the apparently quite
different way of exchanging a general state in Fig.~2 is to replace
the cNOT gate on the left of Fig.~2 with an elementary classical
circuit, only slightly more elaborate than that of Fig.~1, that
changes the direct coupling of the cNOT into four couplings, all
acting only through the intermediary of an unaltered ancillary qubit.
\begin{figure} \centerline{\psfig{file=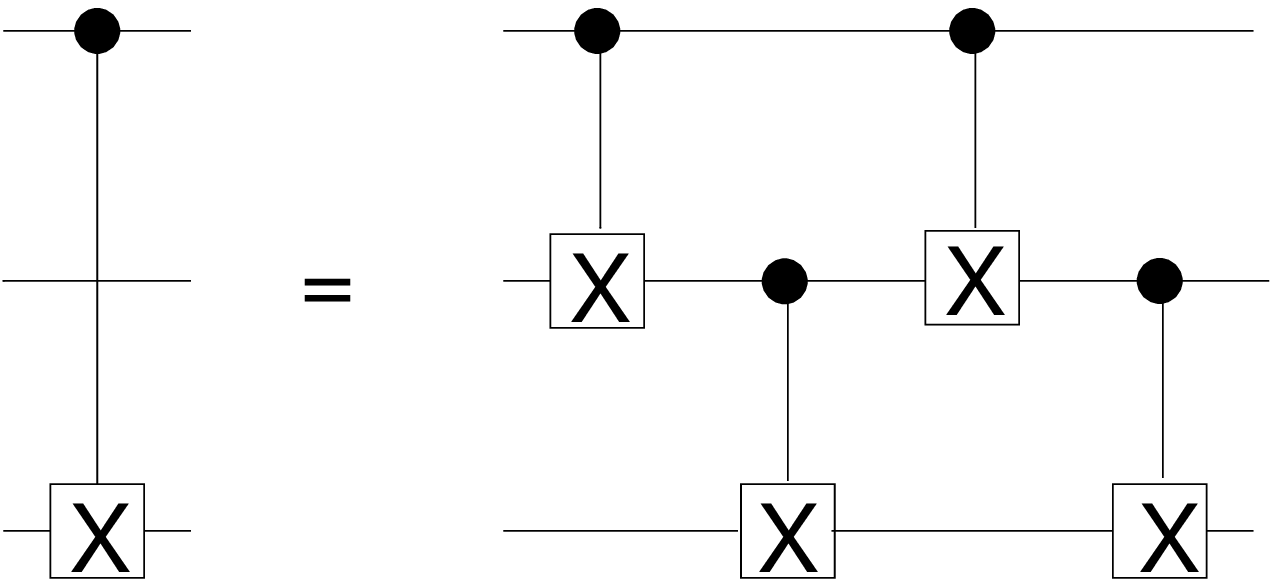,width=3truein}}
\caption{}
\end{figure} \noindent To confirm this identity note that the four gates on the right act as follows on
the eight computational basis states $\k x\k y\k z$ (with $\k x$ the
input state on the top left, $\k z$ on the bottom, and $\k y$ in the
middle) \cite{ft:fig3}:  \begin{equation} \matrix{\k x\k y\k z &\ra \k x\k{y\+x}\k z
\ra \k x \k {y\+x} \k {z\+y\+x}\hfill\cr &\ra \k x \k y \k {z\+y\+x}
\ra \k x\k y\k {z\+ x}.\hfill\cr} \label{eq:4cnot} \end{equation} Thus
the circuit on the right of Fig.~3 does indeed act as indicated on the
left, performing a cNOT on the qubits associated with the top
(control) and bottom (target) wires, while acting as the identity on
the qubit associated with the middle wire.

Quantum mechanics first appears when we interchange control and target
in the cNOT gate on the right of Fig.~2, using the quantum circuit
identity
\begin{figure}
\centerline{\psfig{file=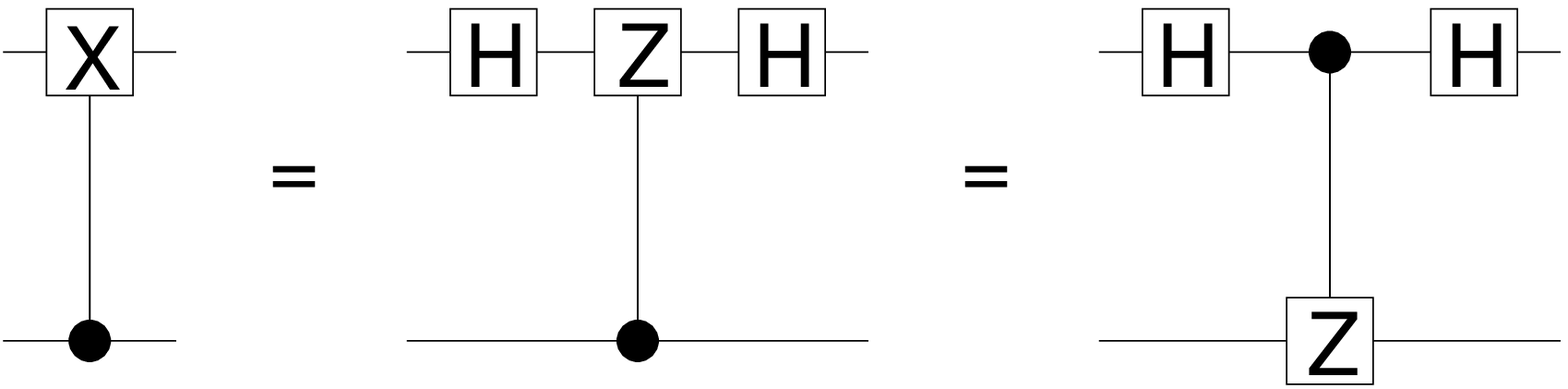,width=3.3truein}} 
\caption{}
\end{figure}

\noindent This follows from the fact that the unitary, self-inverse,
Hadamard operator $\Hc = \c(\sigma_x+\sigma_z)$ takes eigenstates of
$\Xc = \sigma_x$ into eigenstates of $\Zc = \sigma_z$ with
corresponding eigenvalues, and vice-versa:
\begin{equation}
\Hc:\ \ \ \k 0 \leftrightarrow \c(\k0+\k1),\ \ \ \k1 \leftrightarrow
\c(\k0-\k1),
\label{eq:hadamard}
\end{equation}
together with the fact that
controlled-$\Zc$ has the same action regardless of which qubit is the
target and which the control\cite{ft:cZ}.  The utility of this
interchange emerges below.

So if we introduce an ancilla in a state $\k \chi$, to be specified in a
moment, we can replace the two gates in Fig.~2, with the equivalent
circuits of Figs.~3 and 4, to get
\begin{figure}
\centerline{\psfig{file=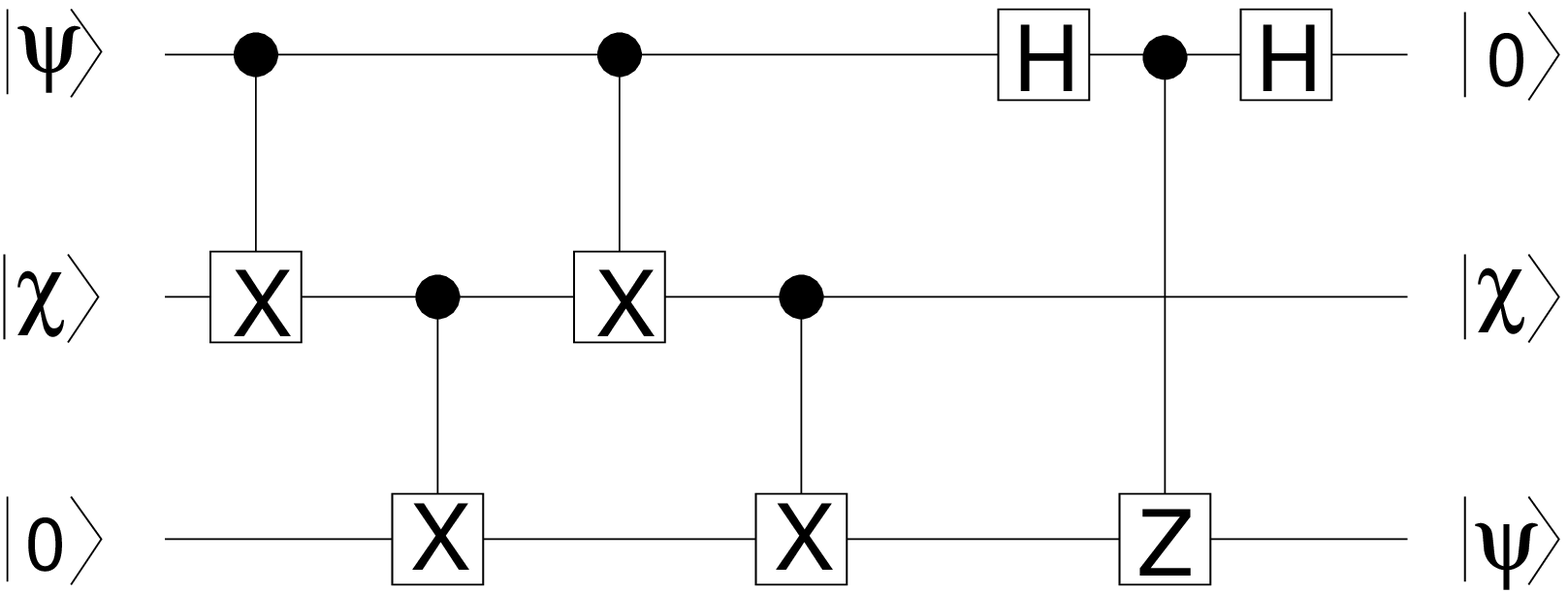,width=3.3truein}} 
\caption{} 
\end{figure} 

\noindent I emphasize that Fig.~5 is merely a cumbersome way of
constructing the classical circuit of Fig.~2, with the direct couping
on the left of Fig.~2 replaced by the four gates on the left, mediated
by an ancillary qubit whose state is unaltered, and the direct
coupling on the right replaced by the three gates on the right, which
by exploiting the quantum-mechanical $\Hc$ gates make it possible to
interchange control and target qubits.

To further convert the circuit of Fig.~5 into teleportation, we must
first eliminate the unacceptable leftmost coupling between the source
and the ancilla. This can be done by taking the state $\k\chi$ of the
ancilla to be $\Hc\k0$, which the magic of quantum mechanics --- this
is the second place where it appears --- allows to be invariant under
NOT.  Because \begin{equation}
\Xc\Hc\k0 = \Hc\k0,\label{eq:XHO}
\end{equation} the leftmost controlled-$\Xc$ in
Fig.~5 always acts as the identity, and can be removed from the
circuit.  So Fig.~5 becomes
\begin{figure}
\centerline{\psfig{file=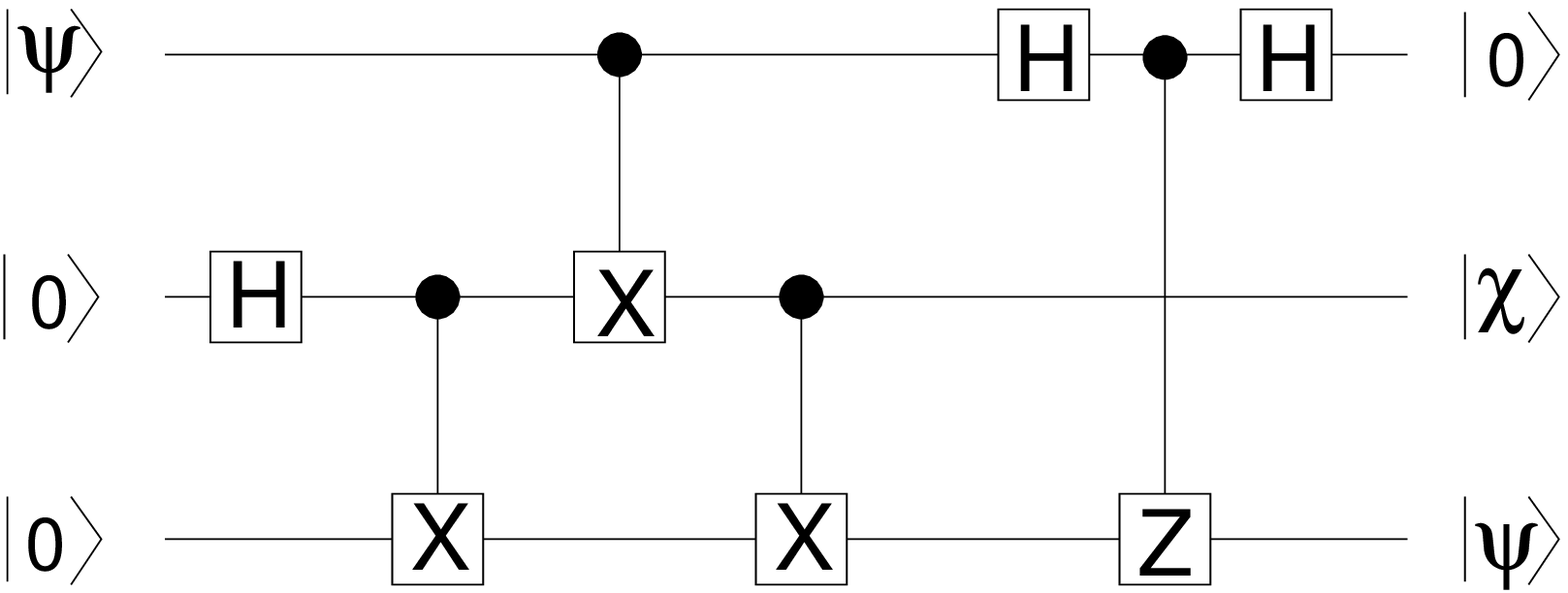,width=3.3truein}} 
\caption{} 
\end{figure} 
To see that Fig.~6 represents quantum teleportation note that we can
also remove the final Hadamard transformation on the upper wire in
Fig.~6, provided we change the final state of the qubit associated
with that wire from $\k 0$ to $\Hc^{-1}\k 0 = \Hc\k 0 = \k\chi$.
Because the remaining Hadamard on the upper wire commutes with the
cNOT that immediately precedes it on the lower two wires, we may also
exchange the order of these two gates.  The result is
\begin{figure}
\centerline{\psfig{file=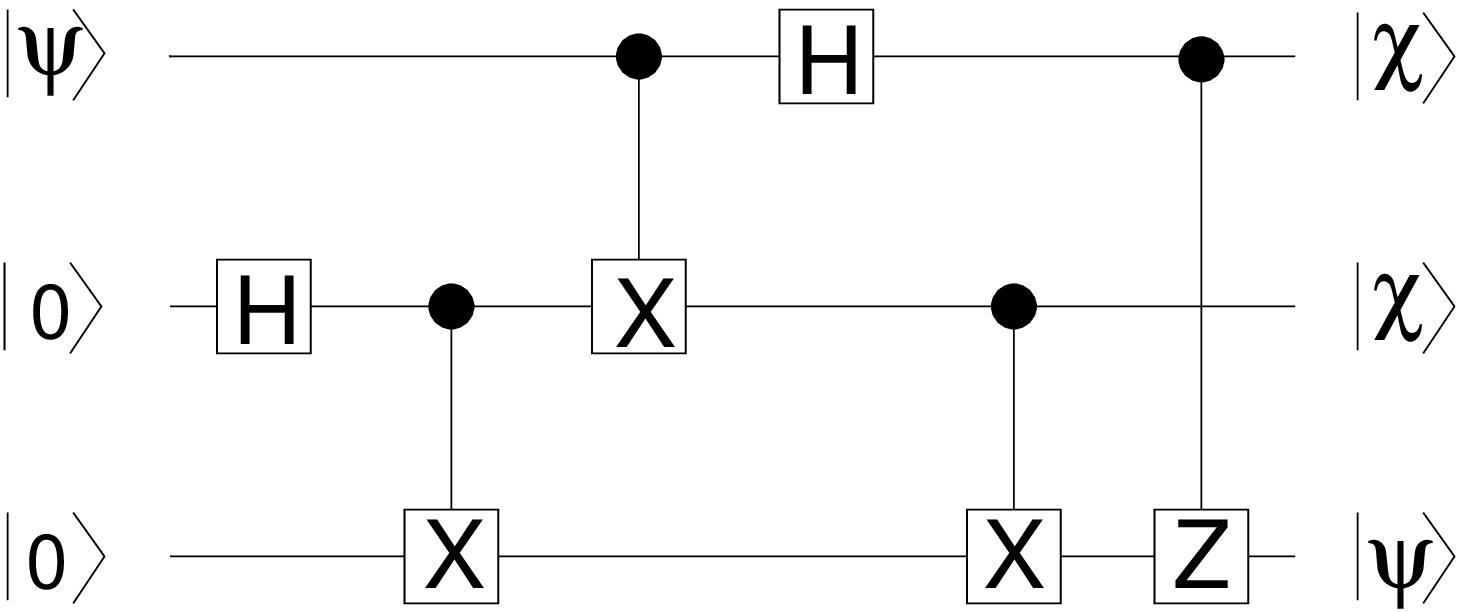,width=3.2truein}} 
\caption{} 
\end{figure} 

\noindent This is precisely the reversible quantum teleportation
circuit described by Brassard, Braunstein, and Cleve
(BBC)\cite{ft:BBC}. We have thus made a direct passage from the
classical circuit of Fig.~2, which requires coupling between source
and destination to swap their states, to the BBC quantum teleportation
circuit of Fig.~7, which, as reviewed below, can be further modified
to remove all remaining coupling.

I repeat BBC's description of the connection between the circuit of
Fig.~7 and teleportation, to indicate what has become of the
couplings originally present in Fig.~2 and to show that the four cNOT
gates arising from the classical expansion in Fig.~3 of the first cNOT
gate in Fig.~2 now play roles in three distinct stages of the quantum
teleportation process!  \cite{ft:repeat} 

The cNOT on the left in Fig.~7, along with the Hadamard gate
immediately to its left, used to eliminate the fourth cNOT from
Fig.~3, serve to turn the state of the ancilla and destination into
the maximally entangled state $\c\bigl(\k 0\k 0+\k 1\k 1\bigr)$.
After these two gates have acted Alice keeps the ancilla and Bob takes
the destination to a faraway place.  Only after that need Alice
acquire the source, in the state $\k\Psi$, which may or may not be
known to her.

The effect of the next cNOT and Hadamard of Fig.~7 on the source and
ancilla, both in Alice's possession, is to transform unitarily the
four mutually orthogonal maximally entangled states of the Bell
basis\cite{ft:bellbasis} into the four computational basis states $\k
x\k y$.  If Alice's two qubits were to be measured in the
computational basis after the action of the first four gates, the
measurement could therefore be viewed as a coherent two-qubit
measurement in the Bell basis, taking place immediately after the
first two gates\cite{ft:bell}.

Such measurements in the computational basis, which are the third and
final place where quantum mechanics enters the process, can be
introduced, though initially at the wrong stage of the process, by
noting that in the final state on the right of Fig.~7 Alice's two
qubits are each in the pure state $\k\chi$, completely disentangled
from Bob's.  As a result, the state of Bob's qubit is entirely
unaffected if Alice measures each of her qubits.  So we can safely add
two measurements to Fig.~7 without disrupting the transfer of $\k\psi$
from Alice's qubit to Bob's:
\begin{figure}
\centerline{\psfig{file=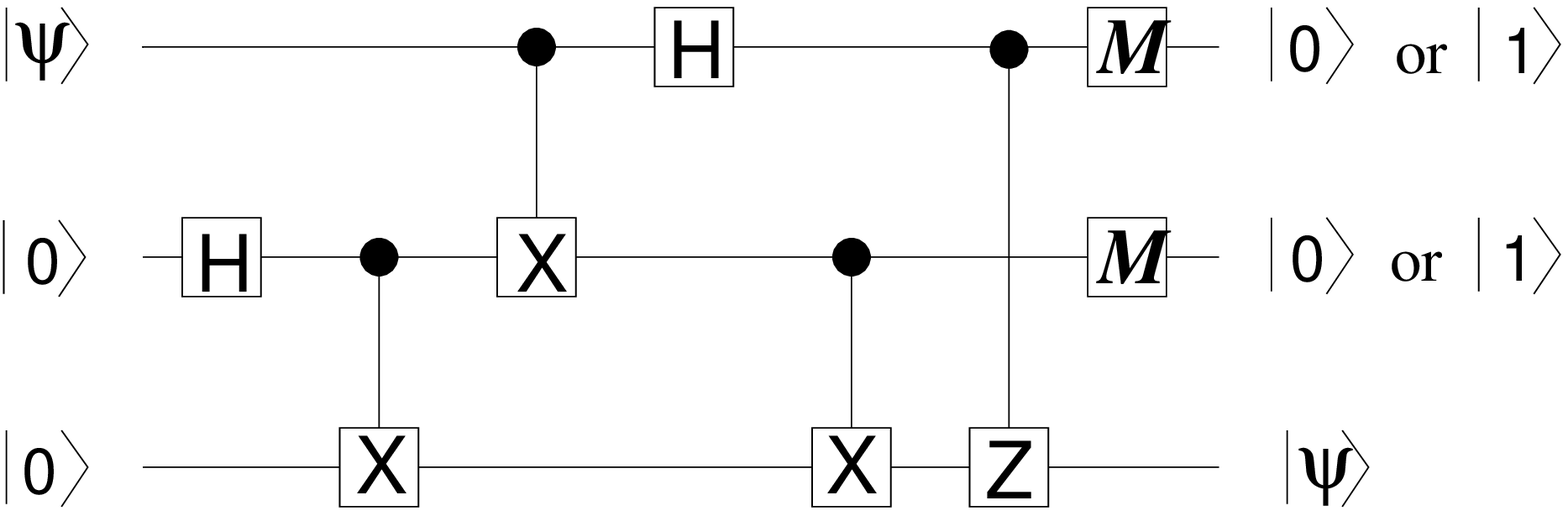,width=3.4truein}} 
\caption{} 
\end{figure} 

Not only do these measurements occur too late in the process, but
there also remain in Fig.~8 two other interactions between Alice's
qubit or her ancilla and Bob's, besides the cNOT gate that originally
entangles her ancilla with his destination.  The controlled-$\Zc$ on
the right comes directly from the controlled-$\Xc$ on the right of
Fig.~2, and the controlled-$\Xc$ immediately preceding it comes from
the last of the four controlled-$\Xc$ gates on the right of Fig.~3.
Both these interactions can be replaced by classical communication of
measurement results from Alice to Bob, by moving the measurements to
the earlier stage of the process mentioned above, which it is possible
to do for the following reason:

Quite generally the effect of a controlled unitary operation on any
number of qubits followed by a measurement of the control qubit is
unaltered if the measurement of the control qubit precedes the
controlled operation\cite{ft:griffiths}:

\begin{figure}
\centerline{\psfig{file=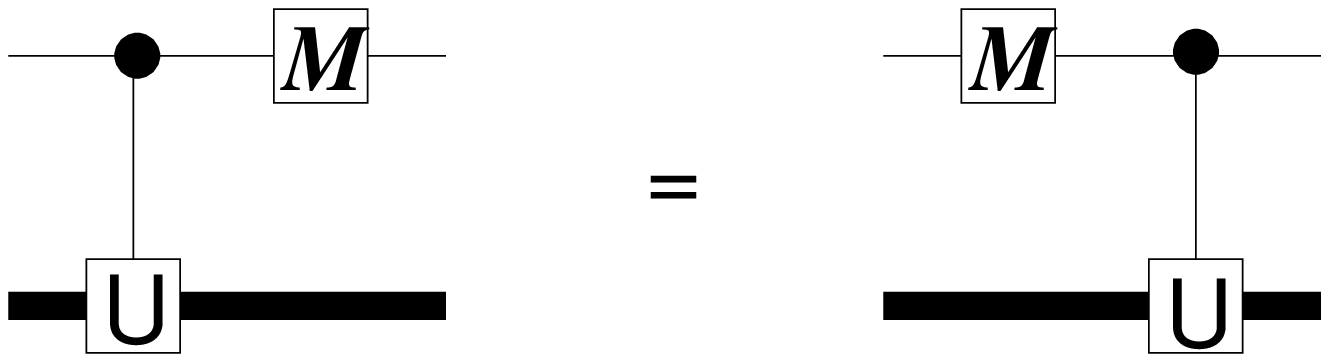,width=3truein}}
\caption{} 
\end{figure}

\noindent Here the heavy horizontal wire represents $N$
additional qubits, and $\Uc$ represents a unitary transformation
acting on any or all of those qubits, controlled by the single qubit
represented by the light wire.  

The measurement and the controlled-unitary operation commute because
an arbitrary input state $\k\Psi$ of the $N+1$ qubits is necessarily
of the form 
\begin{equation} \k\Psi = a\k0\k{\Phi_0} + b\k1\k{\Phi_1}
\label{eq:psi} 
\end{equation} 
where $|a|^2 + |b|^2 = 1$, $\k0$ and
$\k1$ are computational basis states of the control qubit, and
$\k{\Phi_0}$ and $\k{\Phi_1}$ are normalized (but in general
non-orthogonal) states of the other $N$ qubits.  An immediate
measurement on the control qubit takes $\k\Psi$ into $\k0\k{\Phi_0}$
with probability $|a|^2$, or into $\k1\k{\Phi_1}$ with probability
$|b|^2$y\cite{ft:born}. In the first case subsequent application of a
controlled-$\Uc$ has no further effect; in the second case it produces
the state $\k1\,\Uc\k{\Phi_1}$.  

On the other hand an immediate application of the controlled-$\Uc$
operation takes $\k\Psi$ into
\begin{equation} a\k0\k{\Phi_0} + b\k1\,\Uc\k{\Phi_1} \label{eq:uphi}
\end{equation} and a subsequent measurement of the control qubit takes
this state into $\k0\k{\Phi_0}$ with probability $|a|^2$, or
$\k1\,\Uc\k{\Phi_1}$ with probability $|b|^2$.  Thus the two output
states are the same and occur with the same probabilities, regardless
of the order in which the measurement and controlled-$\Uc$ are
performed.

Fig.~9 allows Fig.~8 to be rewritten as
\begin{figure}
\centerline{\psfig{file=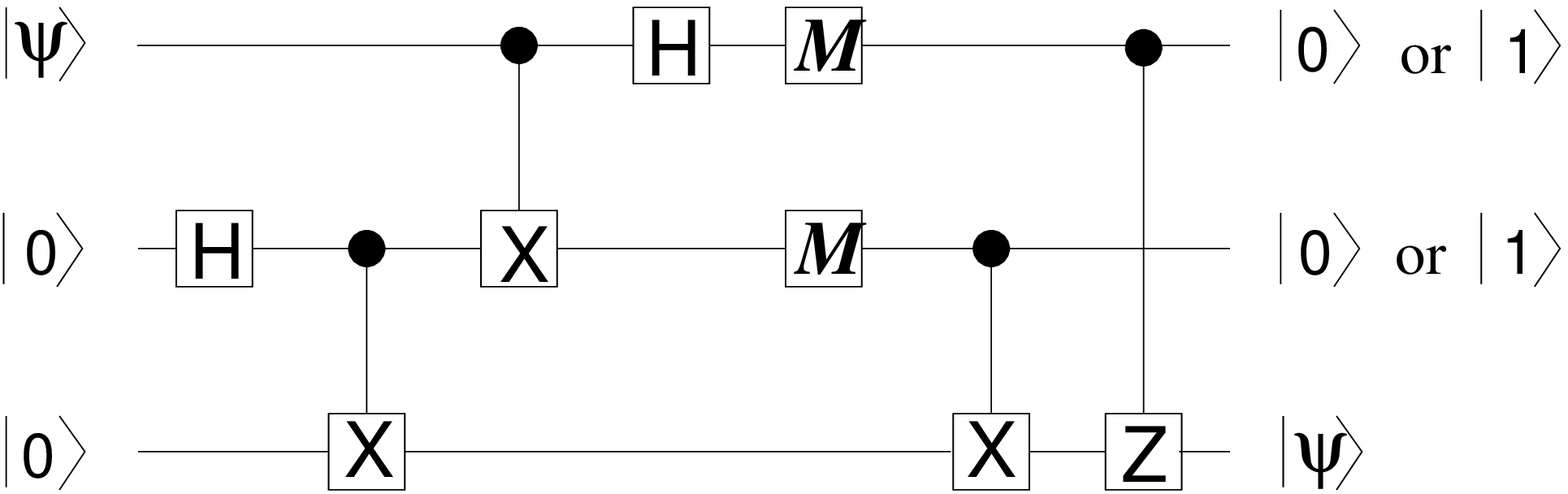,width=3.4truein}} 
\caption{} 
\end{figure} 
\noindent which shifts the actual measurements to the position of the
hypothetical measurements mentioned above.  Since the controlled-$\Xc$
or controlled-$\Zc$ in Fig.~10 now follow a measurement of the control
bit, their action is identical to applying the $\Xc$ or $\Zc$ to the
target qubit if and only if the outcome of the corresponding
measurement is 1; i.e. the controlled operation can be executed
locally by Bob depending on what Alice tells him about the outcomes of
the two measurements she made on her own qubits.

To summarize, we can look at the teleportation protocol of Fig.~10,
and ask what became of the original three couplings in the general
classical state-swapping protocol of Fig.~1.  The coupling on the left
of Fig.~1 vanished by virtue of the initial choice $\k 0$ for the
state of the destination (bottom wire of Fig.~10).  The middle
coupling of Fig.~1 survives in the three cNOT gates coupled to the
ancilla (middle wire) in Fig.~10\cite{ft:first}. Two of the three
cNOT's that remain do indeed provide links from Alice's qubits to the
destination.  But one (on the left of Fig.~10) operates only to create
the initial entanglement of the ancilla with the destination, while
the other (on the right) operates only through Alice's telling Bob,
depending on the result of her measurement on the ancilla, whether or
not to apply the transformation $\Xc$ to the
destination\cite{ft:third}.  The coupling on the right of Fig.~1
survives as the transformation $\Zc$ applied to the destination or not
by Bob depending on what Alice tells him about the result of her
measurement on the source.

So you can take the BBC circuit of Fig.~7 and look back to its
classical ancestry (Fig.~1) or forward to conventional teleportation
(Fig.~10), seeing the same cNOT gates play entirely different roles,
depending on which way you want to view the circuit, rather like an
optical illusion or a piece of kinetic sculpture.  Depending on how
you put the punctuation marks into a sequence of operations, you can
get a process that is either entirely classical or deeply quantum
mechanical.  

\medskip

This view of teleportation as a quantum mechanical deconstruction of a
trivial classical  state-swapping circuit generalizes readily from qubits to
$d$-state systems (``qudits'').  If we are dealing with a $d$-valued
classical register, we can generalize cNOT to the controlled bit
rotation, 

\begin{equation}\cX:\ \ \k x\k y \ra \k x \k{y\+x},\ \ 0 \leq x, y <
d,\label{eq:cX}\end{equation} \noindent where $\+$ now denotes
addition modulo $d$.  This extends by linearity to a unitary operation
on quantum $d$-state systems, which is only self-inverse when $d=2$.
In the general case the inverse is \begin{equation}\cX^\dagger:\ \ \k
x\k y \ra \k x \k{y\ominus x},\ \ 0 \leq x, y <
d,\label{eq:cXadj}\end{equation} where $\ominus$ denotes subtraction
modulo $d$.  The classical circuits of Figs.~2 and 3 thus become

\begin{figure}
\centerline{\psfig{file=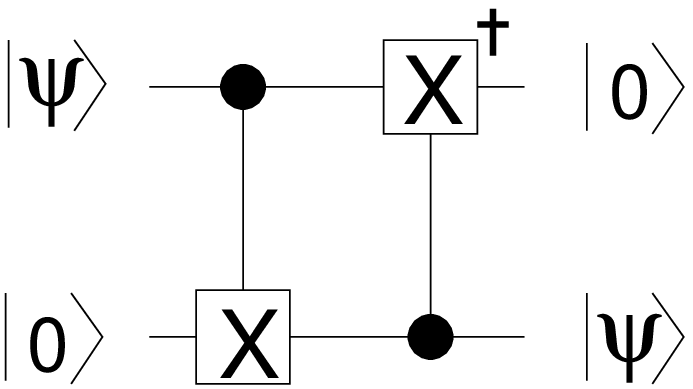,width=1.7truein}} 
\caption{} 
\end{figure} 
\noindent 
and
\begin{figure}
\centerline{\psfig{file=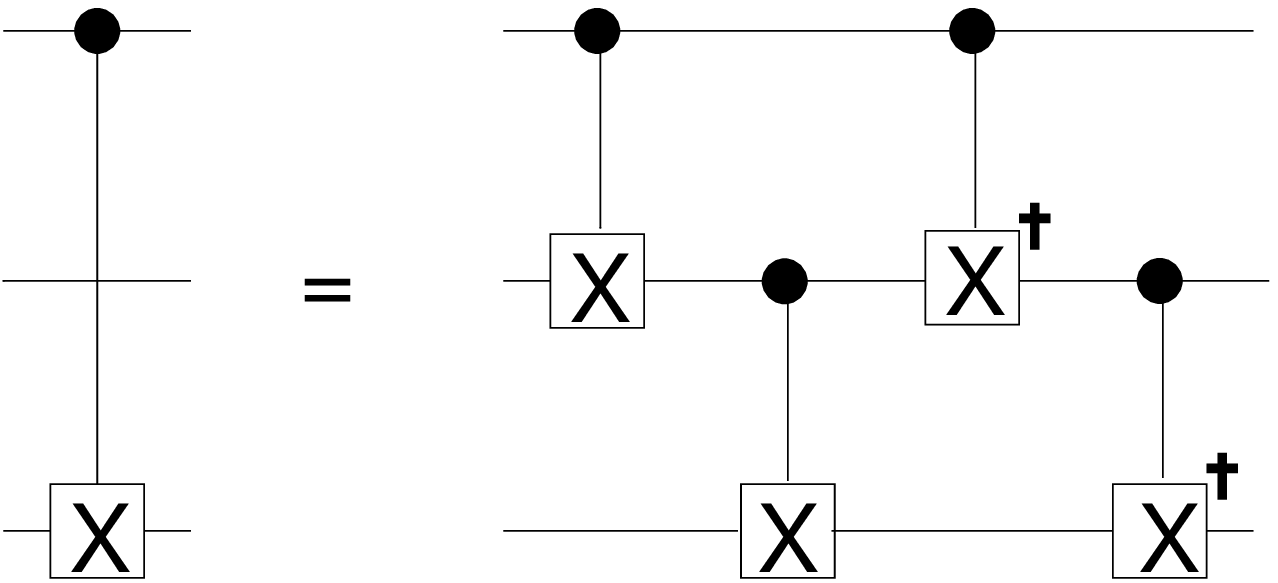,width=3truein}}
\caption{}
\end{figure}

We generalize the Hadamard transformation $\Hc$ on a single qubit to
the quantum Fourier transform $\Fc$ on a single $d$-state system,
\begin{equation}\Fc: \k y \ra {1 \over \sqrt d}\sum_z e^{2\pi i
zy/d}\k z, \label{eq:Fc}\end{equation} and its inverse
\begin{equation}\Fc^\dagger: \k y \ra {1 \over \sqrt d}\sum_z e^{-2\pi
i zy/d}\k z. \label{eq:Fcadj}\end{equation} Note that $\Fc\k0 =
\Fci\k0$ is invariant under an arbitrary bit rotation so that
\begin{equation} (\cX)(\oc\x\Fc)\k\psi\k0 =
\k\psi\k0.\label{eq:inv}\end{equation} A maximally entangled state is
prepared by \begin{equation}(\cX)(\Fc\x\oc)\k0\k0 = {1 \over \sqrt d}
\sum_z\k z\k z. \label{eq:maxentang}\end{equation} (These are the
generalizations of (\ref{eq:epr}) and (\ref{eq:XHO}) from qubits to
qudits.)

An appropriate generalization to $d$-state systems of
controlled-$\sigma_z$ is \begin{equation}\cZ: \k x \k y \ra e^{-2\pi
ixy/d}\k x\k y, \label{eq:cZ}\end{equation} which remains symmetric in control
and target qubits and has the inverse \begin{equation}\cZ^\dagger : \k x
\k y \ra e^{2\pi ixy/d}\k x\k y. \label{eq:cZadj}\end{equation}

In the above definitions of $\cX, \cXi, \cZ, \cZi$ the state on the
left is the control, and the state on the right, the target.  More
generally, in the relations below, let $(\cX)_{ij}$ denote a $\cX$
operation in which state $i$ is the control and state $j$, the target,
and let $(\Fc)_i$ denote a Fourier transform acting on state $i$.

\noindent One easily verifies that \begin{equation} (\cX)_{12}(\Fc)_2 =
(\Fc)_2 (\cZ)_{12}\label{eq:XZ}
\end{equation} 
and
therefore \begin{equation}\cX_{12} =
(\Fc)_2(\cZ)_{12}(\Fci)_2,\label{eq:FZF}
\end{equation}
so
\begin{equation}(\cXi)_{12} = (\Fc)_2(\cZi)_{12}(\Fci)_2 =
(\Fc)_2(\cZi)_{21}(\Fci)_2, \label{eq:ZZ}\end{equation} which has the circuit
representation (the generalization of Fig.~4)\cite{ft:firmly}: 

\begin{figure}
\centerline{\psfig{file=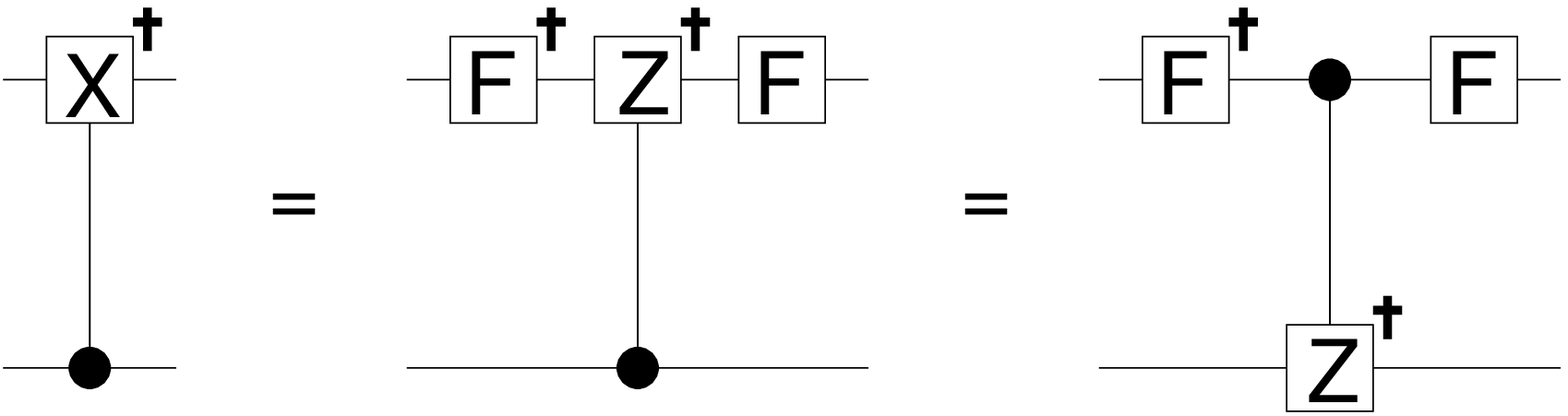,width=3.4truein}} 
\caption{}
\end{figure}

Therefore, following the same sequence of expansions as in the case of
2-state systems, we arrive at the generalization of the BBC circuit of
Fig. 7: \begin{figure}
\centerline{\psfig{file=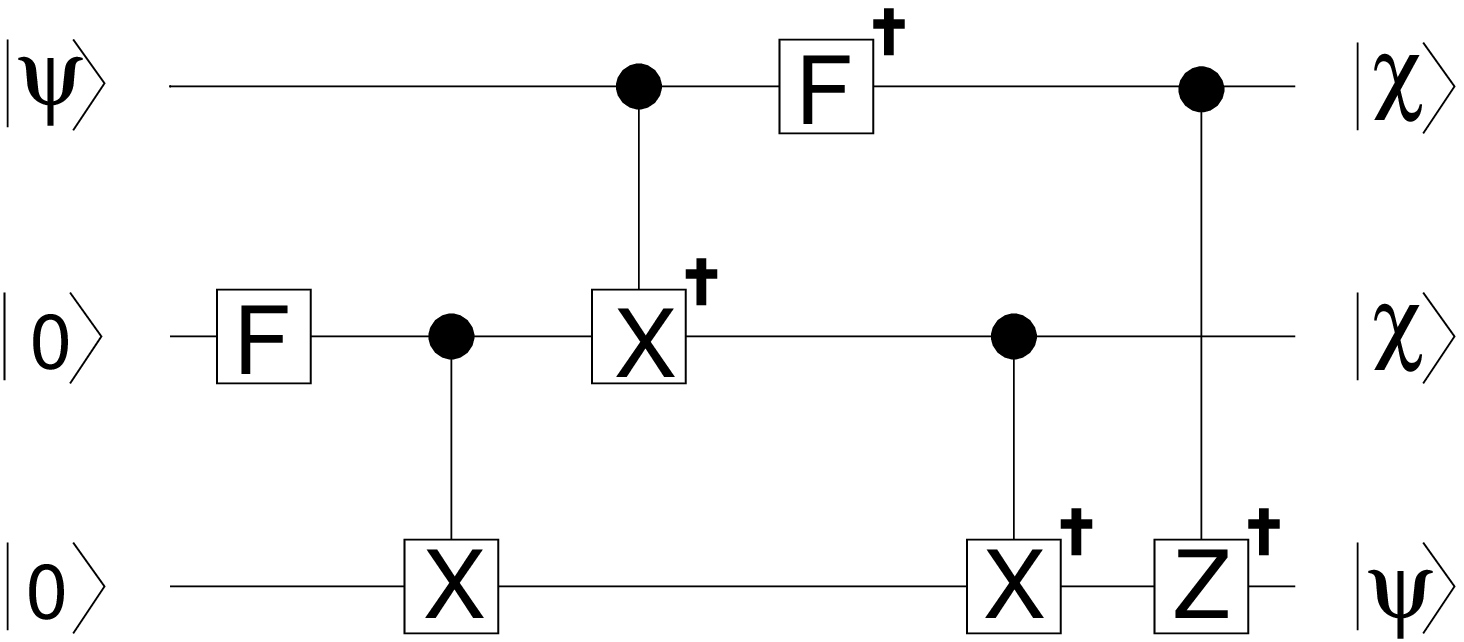,width=3.2truein}} \caption{}
\end{figure} \noindent where \begin{equation}\k\chi = \Fc\k 0 = \Fci\k
0.\label{eq:chid}\end{equation} One can go from this to the
generalization of Fig.~10 \begin{figure}
\centerline{\psfig{file=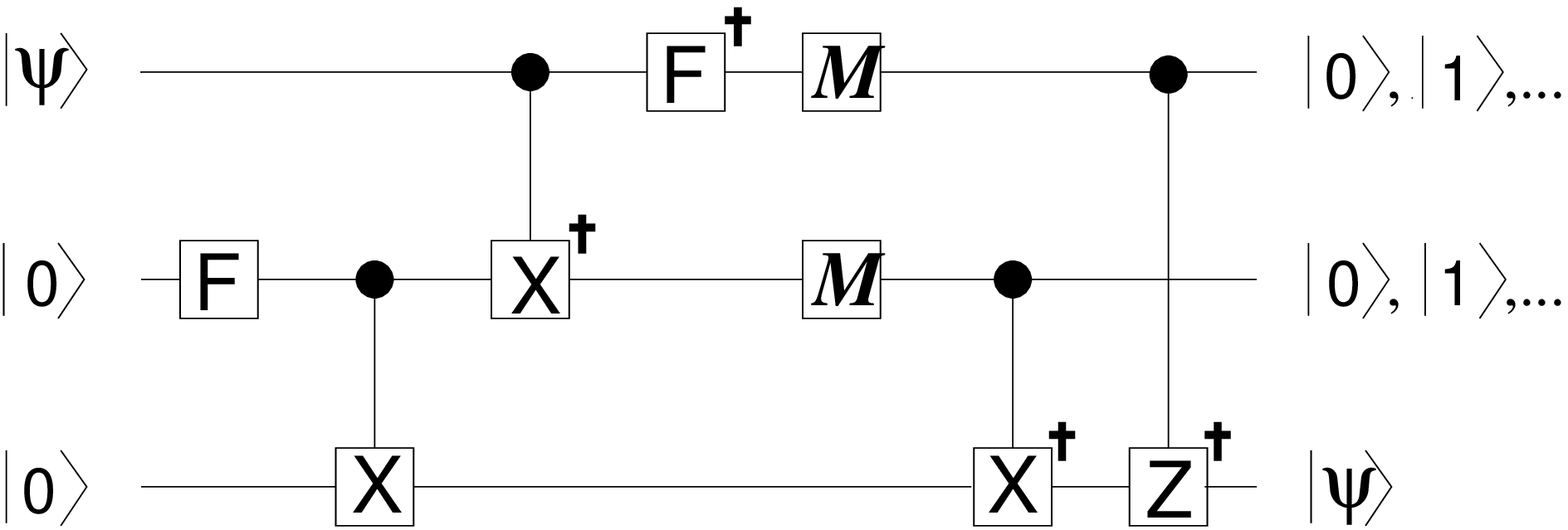,width=3.4truein}} \caption{}
\end{figure} \noindent since the remark \cite{ft:griffiths}, that
measurement of several control qubits commutes with multi-qubit
controlled operations, applies equally well to $d$ state systems even
when $d$ is not a power of 2.

The teleportation circuit of Fig.~15 for $d$-state systems neatly
encapsulates the protocol for teleporting $d$-state systems spelled
out in the original teleportation paper [1], along with its relation to
the protocol of Fig.~10 for teleporting qubits.

I thank Gilles Brassard and Igor Devetak for useful comments on an
earlier version of this essay, and Chris Fuchs for asking why I found
it interesting.  This work is supported by the National Science
Foundation, Grants PHY9722065 and PHY0098429.


\begin{references}
\bibitem{ft:cnot} The unitary cNOT gate operates on the computational
basis --- i.e. the basis of classically meaningful states, $\k0\k0,\
\k0\k1,\ \k1\k0,\ \k1\k1$ ---  as the identity if the state of the control
qubit (indicated by a black dot in Fig.~1) is $\k0$, and flips the
state ($\k0 \leftrightarrow \k1$) of the target qubit (indicated by
the boxed $\Xc$ in Fig.~1) if the state of the control qubit is $\k1$.
\bibitem{ft:classical}
I call a quantum circuit classical if
it is classically meaningful when restricted to classically meaningful
states --- i.e. if every unitary gate takes computational-basis states
into other computational basis states without introducing
superpositions or phases. Because the circuit of Fig.~1 exchanges
computational-basis states and acts linearly on superpositions of
inputs, it also, of course, exchanges arbitrary quantum states.
\bibitem{ft:fig3} Alternatively one can note, in the computational
basis, that if the state of the top wire is $\k0$ then neither of the
NOT operations acts on the middle wire so the two lower self-inverse
cNOT operations act in direct succession, giving the identity.  But if
the state of the top wire is $\k1$ then both NOT operations act on the
middle wire, leaving its state unaltered, and ensuring that exactly
one cNOT operation acts on the lower wire regardless of that state.
\bibitem{ft:repeat}  It is also necessary to retrace this
familiar ground to confirm that it supports the generalization
to $d$-state systems described at the end of this note.
\bibitem{ft:bellbasis}.  The Bell-basis states are $\c\bigl(\k 0\k
0\pm\k 1\k 1\bigr)$ and $\c\bigl(\k 0\k 1\pm\k 1\k 0\bigr)$.  It is
easiest to see that the cNOT and Hadamard gates have this affect by
looking at the inverse transformation.
\bibitem{ft:bell}Conventional expositions of teleportation do indeed
expand the state of Alice's two qubits in the Bell basis after the
entangled pair is formed, having her then make a coherent two-qubit
measurement in that basis.  But it is simpler analytically when
algebraically tracing the progress of a general $\k\psi$ through the
protocol, as well as more straightforward to implement physically, to
take seriously the circuit of BBC, letting Alice explicitly apply the
next cNOT and Hadamard and follow this by independent qubit
measurements in the ordinary computational basis.  As BBC note, there
is no need to mention the Bell basis at all.
\bibitem{ft:born}This extension of Born's probability rule to cases in
which only a subsystem is measured, which is crucial in quantum
computation, receives surprisingly little explicit attention in most
textbook introductions to quantum mechanics.
\bibitem{ft:first}The very first of the four cNOT gates coming from
the expansion in Fig.~3 of the middle coupling of Fig.~1 was crucially
rendered unnecessary by the initial choice $\Hc\k0$ for the state of
the ancilla.
\bibitem{ft:third}The remaining cNOT in Fig.~10 links Alice's qubit
only to her ancilla.  It can be viewed, if one wishes, as a part of
the process of ``measurement in the Bell basis.''
\bibitem{ft:firmly}Note the unfortunate but firmly entrenched
convention that in circuit diagrams operations on the left act first
while in equations operations on the right act first.
\end{references}
\end{document}